\documentclass[a4paper,11pt,notitlepage]{article}

\usepackage[utf8x]{inputenc}

\title{\textbf{Displacement Damage dose and DLTS Analyses on Triple and Single Junction solar cells irradiated with electrons and protons}}
\author{Carsten Baur$^{5}$, Roberta Campesato$^{1}$, Mariacristina Casale$^{1}$, Massimo Gervasi$^{2,3}$,\\ Enos Gombia$^{4}$, Erminio Greco$^{1}$, Aldo Kingma$^{4}$, Pier Giorgio  Rancoita$^{2}$,\\ Davide Rozza$^{2,3}$, Mauro Tacconi$^{2,3}$.}
\date{}

\usepackage[utf8x]{inputenc}
\usepackage{amsmath}
\usepackage{amsfonts}
\usepackage{amssymb}
\usepackage{xcolor,graphicx}
\usepackage[top=2.3cm,bottom=2.3cm,left=2.3cm,right=2.3cm]{geometry}

\begin{document}
\maketitle
\begin{center}
$^1$ \textit{CESI, via Rubattino 54, I-20134 Milan, Italy}\\
$^2$ \textit{INFN Sezione di Milano Bicocca, I-20126 Milan, Italy}\\
$^3$ \textit{Universit\`a di Milano Bicocca, I-20126 Milan, Italy} \\
$^4$ \textit{IMEM-CNR Institute, Parco Area delle Scienze 37/A, 43124 Parma, Italy}\\
$^5$ \textit{ESA/ ESTEC, Keplerlaan 1, 2201 AZ Noordwijk, The Netherlands}\\
\vspace{0.5cm}
Abstract accepted for poster session at \\2017 IEEE Nuclear and Space Radiation Effects Conference, \\July 17-21, New Orleans.
\end{center}

\begin{abstract}
Space solar cells radiation hardness is of fundamental  importance in view of the future missions towards harsh radiation environment (like e.g. missions to Jupiter) and for the new spacecraft using electrical propulsion. In this paper we report the radiation data for triple junction (TJ) solar cells and related component cells. Triple junction solar cells, InGaP top cells and GaAs middle cells degrade after electron radiation as expected. With proton irradiation, a high spread in the remaining factors was observed, especially for the TJ and bottom cells.   Very surprising was the germanium bottom junction that showed very high degradation after protons whereas it is quite stable against electrons. Radiation results have been analyzed by means of the Displacement Damage Dose method and DLTS spectroscopy.
\end{abstract}

\section{Introduction}
In the last 10 years spacecraft are mainly powered by Triple junction solar cells based on III-V compound semiconductors. These solar cells are characterized by 30\% conversion efficiency and a very high  radiation hardness that allow to extend mission lifetime and to use electric propulsion.\\
The radiation analysis of solar cells is very important to predict the End Of Life (EOL) performances of the solar arrays.\\
The test of the solar cell radiation hardness is conducted on Earth by irradiating the solar cells using protons and electrons at different energies.\\
Of course, it is not possible to cover the full spectrum of charged particle energies in space, therefore the experiments on Earth are generally limited to a few energies for electrons and protons.\\
The evaluation of the radiation hardness of the solar cells is performed by means of two methods:
\begin{itemize}
\item The Equivalent Fluence method from JPL\cite{Anspaugh}\;;
\item The Displacement Damage Dose (DDD) from NRL \cite{Messengers}\;.
\end{itemize}
The JPL method is historically the first implemented and used by the space actors; this method uses empirically determined relative damage coefficients (RDCs) and has the advantage of being immediately understandable but, to generate a good estimation, several particle energies shall be used (at least 8 for protons and 3 for electrons).\\
The DDD method  uses calculated values of non-ionizing energy loss (NIEL); the main advantage of this method is that it requires only 1 energy for protons and 2 energies for electrons in order to predict the EOL behavior of solar cells.\\
In this paper, we will present the results of electron and proton irradiation on triple junction solar cells and related component cells manufactured by CESI.\\
The analysis of the radiation hardness is conducted by the DDD method using an innovative approach for the NIEL calculation.

\section{Triple junction solar cells and component cells}
InGaP/InGaAs/Ge triple junction (TJ) solar cells and related component cells  with a size of 4 cm$^2$ and AM0 efficiency class 30\% (CTJ30), have been manufactured and qualified following the ESA ECSS E ST20-08C standard.\\
These solar cells have been developed on a large MOCVD epitaxial reactor, VEECO E450G, suitable to simultaneously process up to thirteen 4-inch wafers per run.\\
The improvement in conversion efficiency was obtained by introducing Quantum Structures, mainly Bragg Reflectors, inside the solar cell stack and by a fine tuning of the electrical field inside the solar cell active regions\cite{Gori}\;.\\
These solar cells have standard thickness of 140 $\mu$m and have been used  on several satellites since 2013.\\
\begin{figure}[ht!]
\begin{center}
 \includegraphics[width=0.24\textwidth]{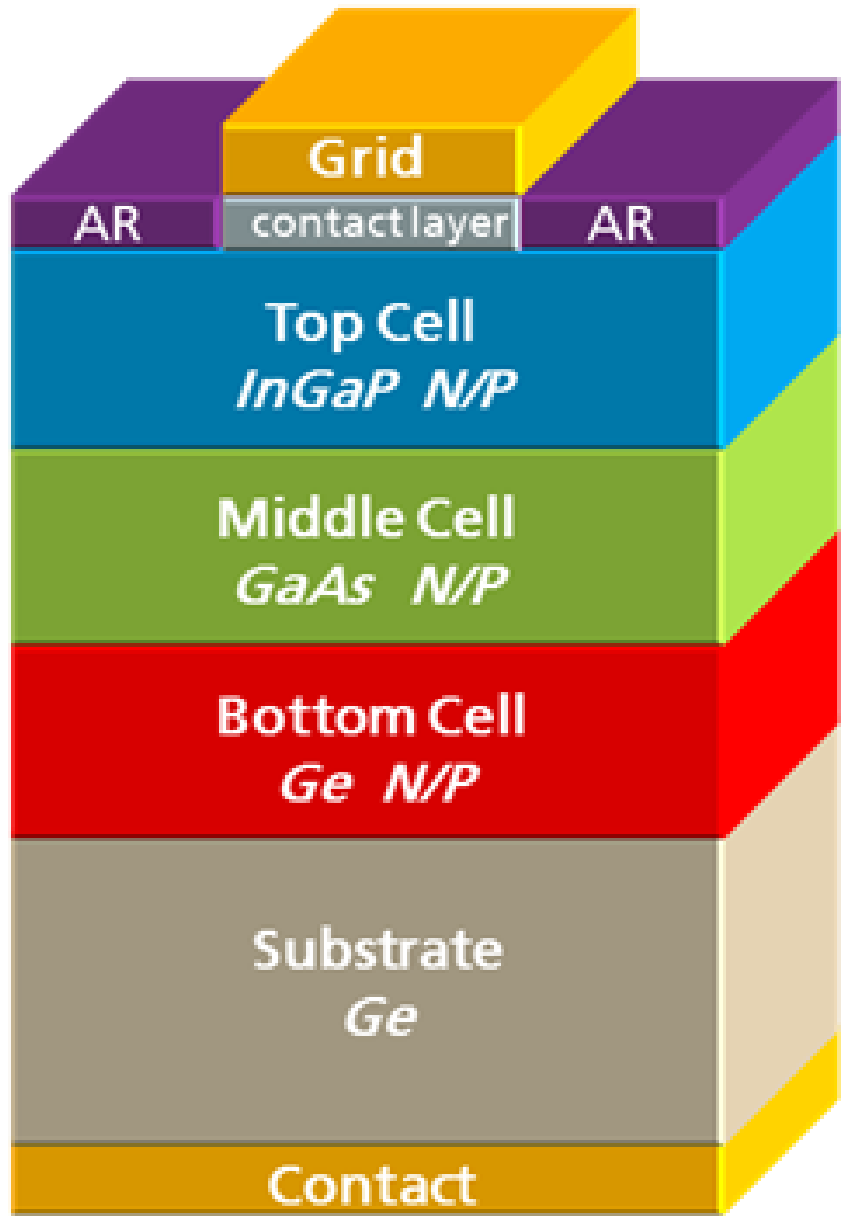}
 \includegraphics[width=0.22\textwidth]{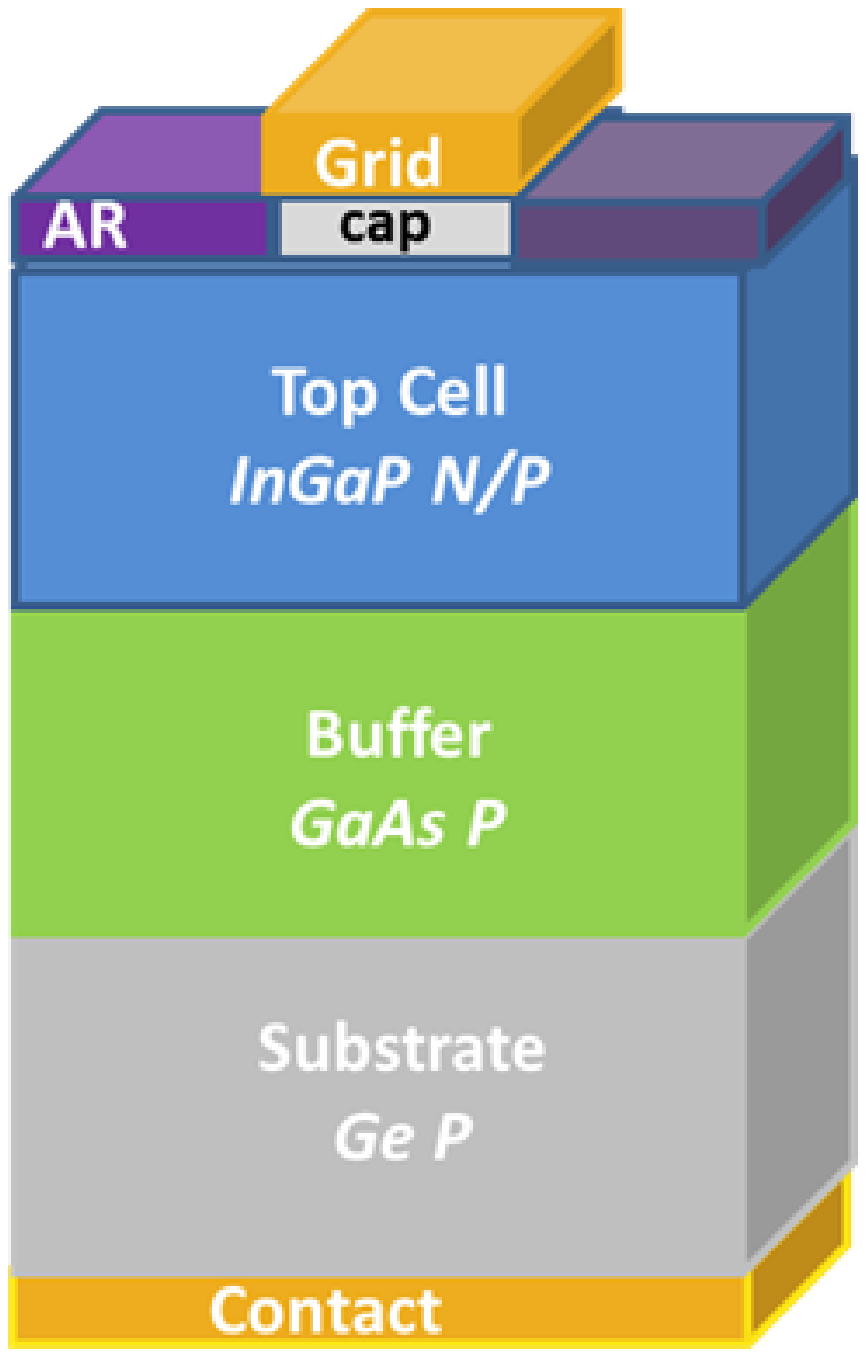}
 \includegraphics[width=0.22\textwidth]{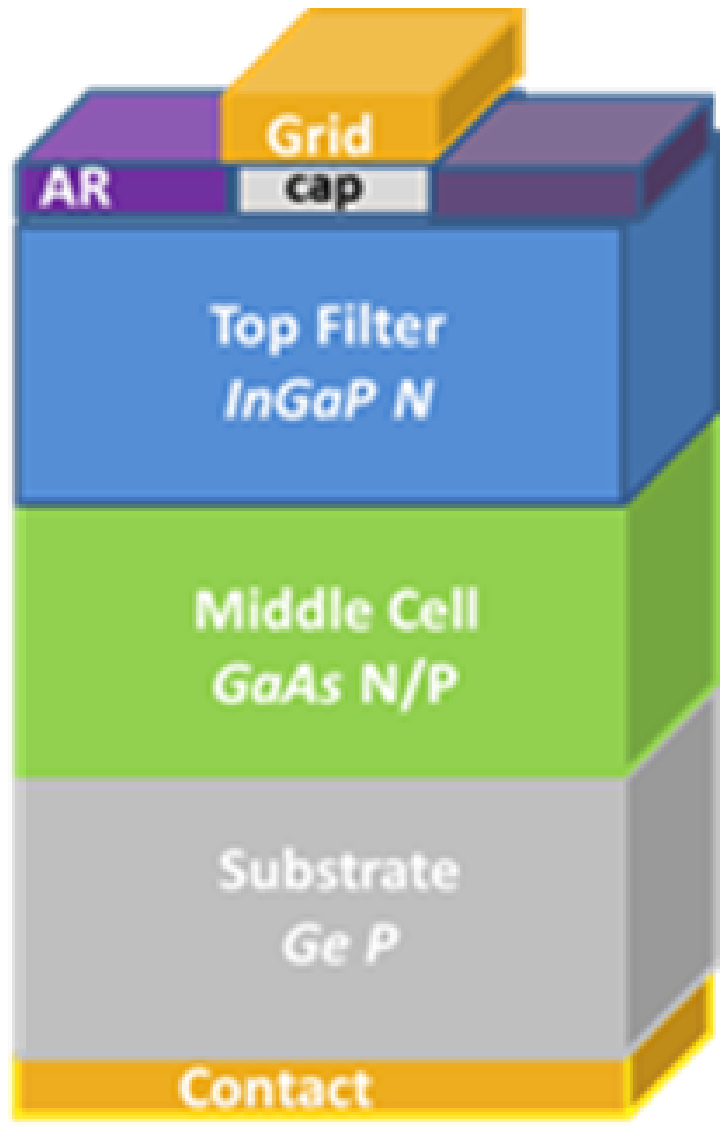}
 \includegraphics[width=0.22\textwidth]{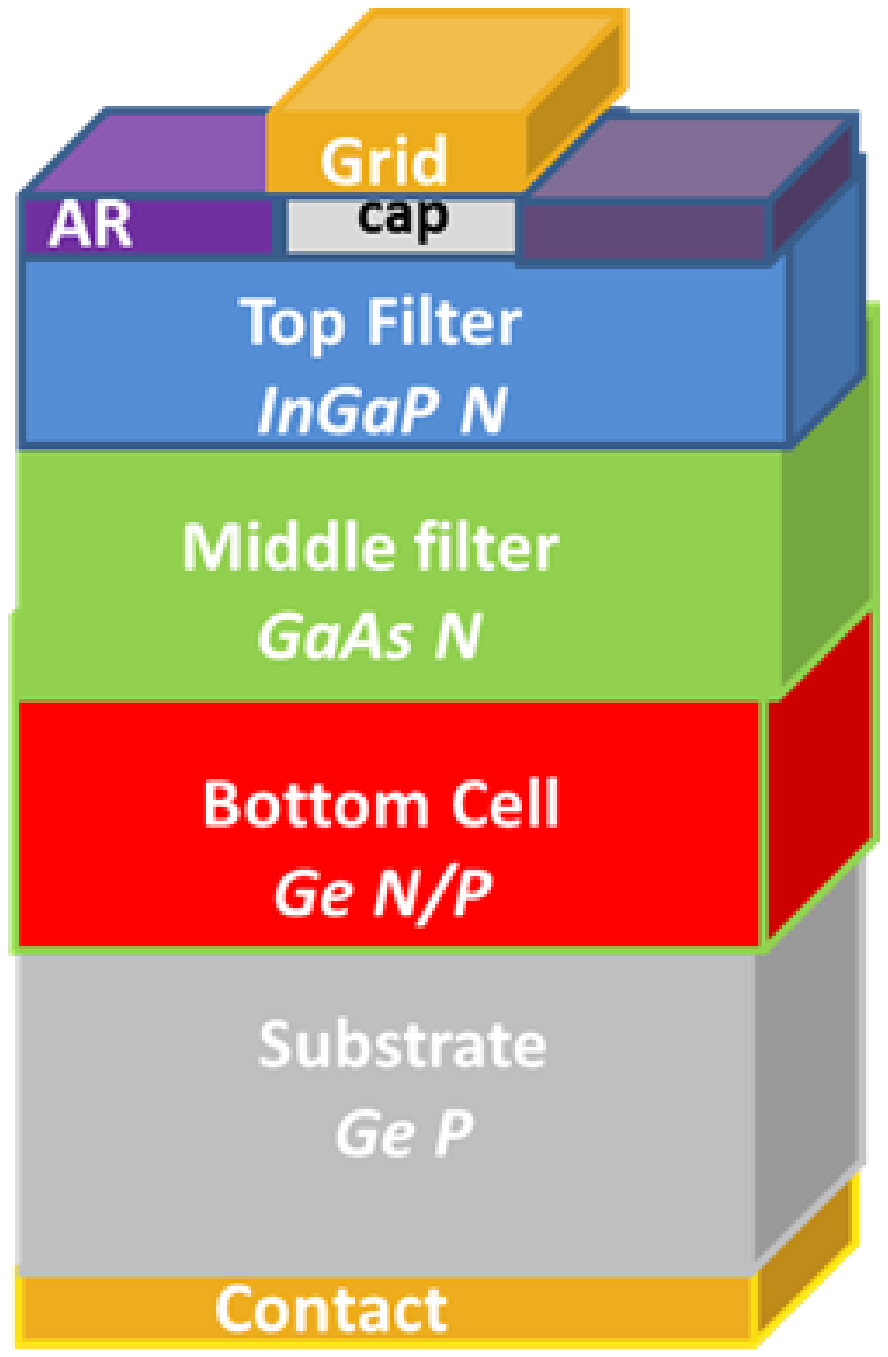}
 \caption{Scheme of a TJ cell(a), top(b), mid(c), bot(d).}
 \label{fig:SchemeTJ}
\end{center}
\end{figure}
The basic structure of the solar cells is reported in figure \ref{fig:SchemeTJ}. The TJ solar cell  is composed by a germanium bottom junction obtained by diffusion into the germanium P-type substrate, a Middle junction of (In)GaAs, whose energy gap is around 1.38 eV and a top junction of InGaP with energy gap 1.85 eV. Component cells are single-junction cells which shall be an electrical and optical representation of the subcells inside the TJ cell. Therefore, to manufacture them, special attention was put to reproduce the optical thicknesses of all the upper layers present in the TJ structure. For example, for the Middle component cell, an n-doped InGaP layer with the same thickness as the top cell was added to absorb the blue portion of the spectrum.

\section{Experimental irradiation results}
TJ solar cells and  component cells have been irradiated with protons and electrons at different fluences. Figure 2 shows the remaining power factors for TJ solar cells obtained for self annealing (solar cells kept in a dry box for 1 month before EOL measurement) and after annealing (8 hours in AM0, 60$^{\circ}$C).\\
An important observation is related to the high spread of results in remaining factors obtained when solar cells are irradiated with low energy protons (0.7 MeV).\\
\begin{figure}[ht!]
\begin{center}
 \includegraphics[width=0.8\textwidth]{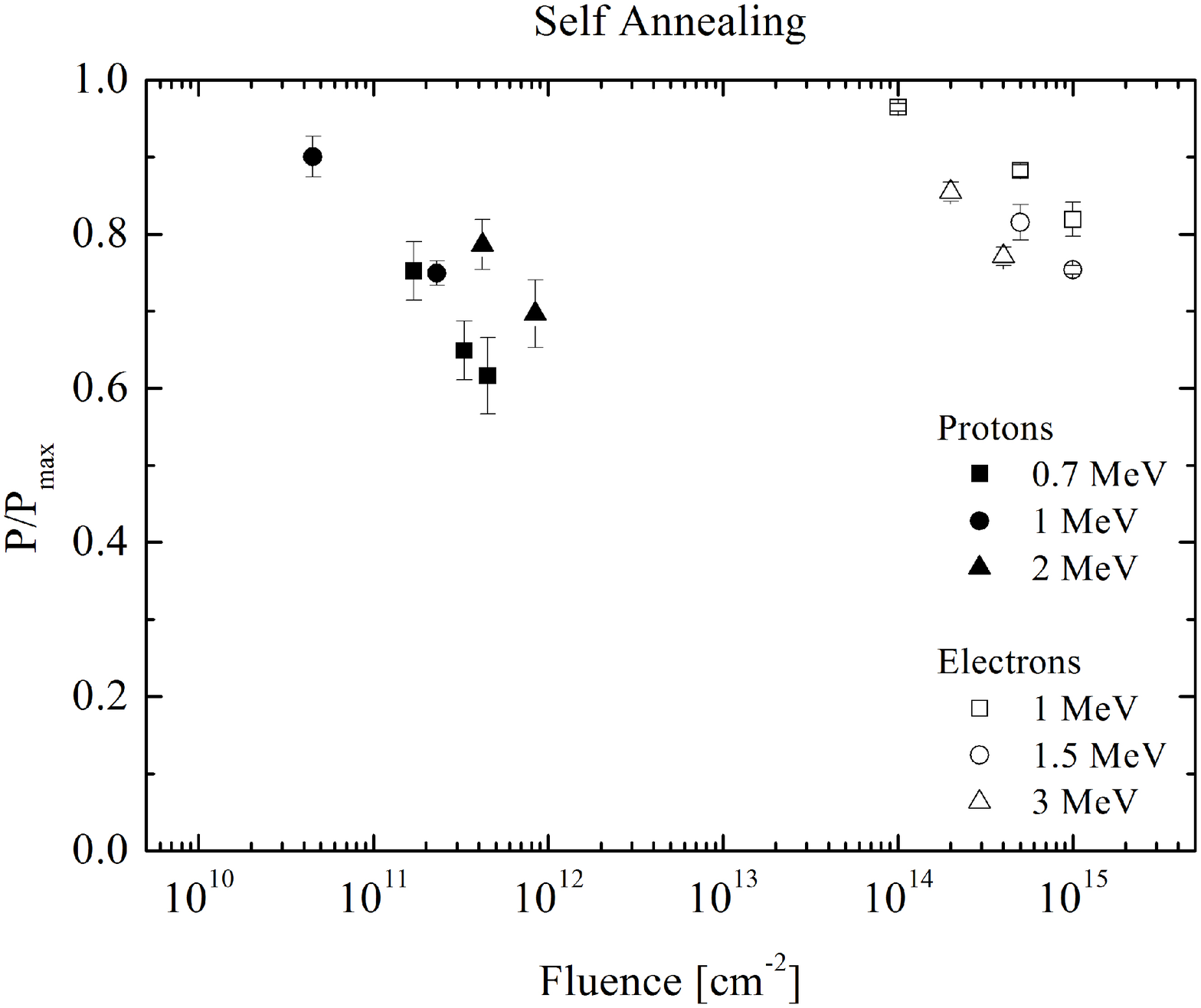}
 \includegraphics[width=0.8\textwidth]{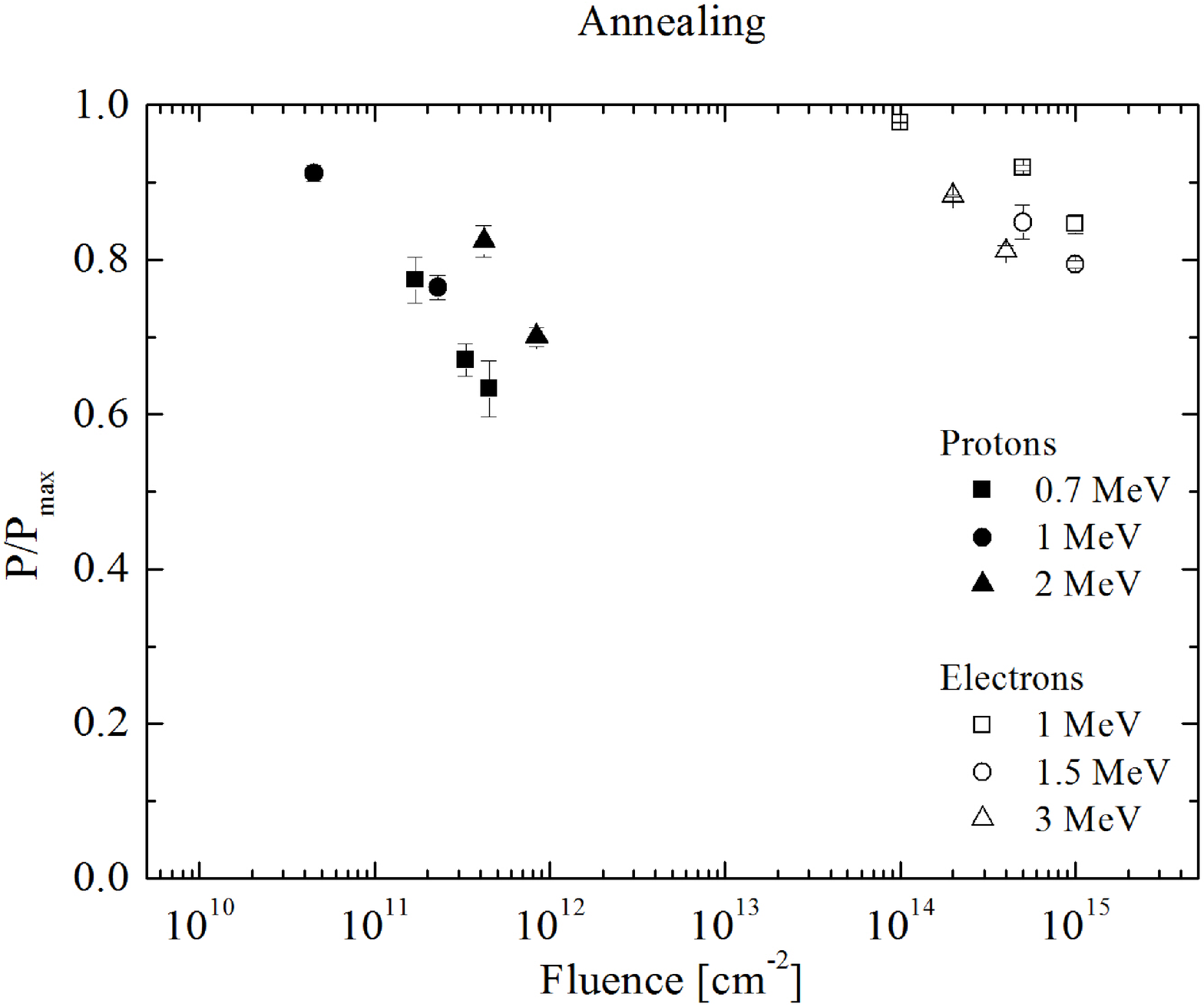}
 \caption{Remaining power factors of TJ solar cells.}
 \label{fig:PowerFactor}
\end{center}
\end{figure}
When component cells EOL behavior was analyzed, it turned out that the weakest and more unstable junction is the bottom one.\\
Germanium is highly resistant against electrons\cite{Baur} but it shows a bad radiation resistance against proton irradiation, especially if low energy protons are concerned.\\
Table 1 shows the main results for TJ and component cells after proton irradiation at 0.7 MeV, fluence $4.5\cdot10^{11}$ p$^+$/cm$^2$.\\
The irradiation was performed by CSNSM on 3 samples for each solar cell type whose area was 4 cm$^2$.\\
The bottom junction is highly degraded after proton irradiation whereas it is highly radiation resistant when irradiated with electrons (table 2). At the irradiation of $4.5\cdot10^{11}$ p$^+$/cm$^2$ with 0.7 MeV protons, the bottom junction has a shunted I-V curve, for this reason the short circuit current of the TJ is limited by the middle junction but probably the bad behavior of the germanium junction could explain the low fill factor observed in TJ solar cells.\\
After annealing, the V$_{oc}$ of Top and Mid cells increases thus improving the V$_{oc}$ of the TJ as expected. The bottom junction seems to recover a portion of the short circuit current (+10\% after annealing) but the shunted I-V curve is still present.\\
\begin{table}[ht!]
\begin{center}
\caption{Electrical performances of TJ, top, mid, bottom (4 cm$^2$ area) before and after irradiation with 0.7 MeV protons, $4.5\cdot10^{11}$ p$^+$/cm$^2$}
{\begin{tabular}{@{}cccccc@{}}
\hline\\[-6pt] & I$_{sc}$[A] & V$_{oc}$[V] & P$_{max}$[W] & F.F. & Eff.[\%]\\[3pt]\hline\\[-6pt]
 & \multicolumn{5}{c}{BOL MEASUREMENT} \\[3pt]\hline\\[-6pt]
TJ & 0.070 & 2.608 & 0.152 & 0.84 & 28.3 \\
Top & 0.071 & 1.365 & 0.082 & 0.85 & 15.03 \\
Mid & 0.077 & 1.012 & 0.065 & 0.84 & 11.91 \\
Bot	& 0.105 & 0.220 & 0.010 & 0.46 & 2.01 \\[3pt]\hline\\[-6pt]
 & \multicolumn{5}{c}{Self annealing} \\[3pt]\hline\\[-6pt]
TJ & 0.056 & 2.207 & 0.094 & 0.75 & 17.12 \\
Top & 0.069 & 1.197 & 0.065 & 0.78 & 11.91 \\
Mid & 0.057 & 0.806 & 0.034 & 0.75 & 6.24 \\
bot & 0.039 & 0.158 & 0.002 & 0.35 & 0.39 \\[3pt]\hline\\[-6pt]
 & \multicolumn{5}{c}{After annealing} \\[3pt]\hline\\[-6pt]
TJ & 0.057 & 2.232 & 0.097 & 0.77 & 17.70 \\
Top & 0.069 & 1.219 & 0.067 & 0.80 & 12.31 \\
Mid & 0.057 & 0.815 & 0.035 & 0.75 & 6.42 \\
Bot & 0.045 & 0.166 & 0.003 & 0.35 & 0.48 \\ \hline
\end{tabular}}
\label{tab1}
\end{center}
\end{table}
\begin{table}[ht!]
\begin{center}
\caption{Remaining factors of bottom cells (4 cm$^2$ area) irradiated with 1 MeV electrons.}
{\begin{tabular}{@{}cccc@{}}
\hline\\[-6pt] 1 MeV e- fluence & RF I$_{sc}$ & RF V$_{oc}$ & RF P$_{max}$\\[3pt]\hline\\[-6pt]
1.00$\cdot10^{14}$ & 0.94 & 0.99 & 0.92 \\
5.00$\cdot10^{14}$ & 0.90 & 0.96 & 0.84 \\
1.00$\cdot10^{15}$ & 0.88 & 0.96 & 0.82 \\ \hline
\end{tabular}}
\label{tab2}
\end{center}
\end{table}
Smaller diodes of 1 mm$^2$ with the same epitaxial structure of middle and top component cells, have been irradiated in order to measure DLTS spectra on them.
In the final paper, all the results after irradiation will be presented.

\section{Niel analysis}
In this section the photovoltaic parameters of the TJ solar cell, and single junction cells are investigated as a function of displacement damage dose ($D^{NIEL}$) which is the product of the particle fluence $\Phi$ and the displacement mass-stopping power ${dE_{\rm de}}/{d\chi}$ (i.e., the so-called NIEL in MeV\,cm$^2$/g), which was calculated by means of the SR (Screened Relativist) treatment\cite{Boschini}\;:
\begin{equation}\label{Eq_DDD}
{D^{NIEL}}\left(E_d\right)=\Phi\frac{dE_{de}}{d\chi},
\end{equation}
with
\begin{equation}\label{Eq_dEdchi}
 \frac{dE_{de}}{d\chi}=\frac{N}{A}\int_{E_d}^{E_R^{max}}{E_RL\left(E_R\right)\frac{d\sigma\left(E,E_R\right)}{dE_R}dE_R},
\end{equation}
where $\chi =x \rho_{\rm A}$, $\rho_{\rm A}$ the absorber density in g/cm$^3$, $N$ is the Avogadro constant; $A$ is the atomic weight of the medium; $E$ is the kinetic energy of the incoming particle; $E_R$ and $E_R^{max}$ are the recoil kinetic energy and the maximum energy transferred to the recoil nucleus respectively; $E_d$ the displacement threshold energy; $L(E_R)$ is the Lindhard partition function; $d\sigma(E,E_R)/dE_R$ is the differential cross section for elastic Coulomb scattering for electrons or protons on nuclei. By inspection of equation \ref{Eq_dEdchi} one can remark that $D^{NIEL}$ depends on the displacement threshold energy $E_d$.
Furthermore, one can note that, for electrons, there is no relevant kinetic energy variation along the path inside the absorber (i.e. the TJ solar cell). On the contrary, such a change occurs for protons. In fact, their actual energy, in each junction, could be estimated by means of SRIM\cite{Ziegler}\;. Thus, the doses were computed for the corresponding proton kinetic energies. For example a proton of 0.7 MeV loses about 30\% of its initial kinetic energy before reaching the center of the middle junction, and 40\% before reaching the center of the bottom junction.\\
In the current study, the $D^{NIEL}$ for the TJ solar cells (see figure 3a) are those evaluated for the middle GaAs cell.\\
In addition, the relative degradation of Pmax, Isc, and Voc obtained after irradiation for the bottom cell exhibit an expected sudden drop. This was already observed and explained in \cite{Baur}\;. Therefore, the three sets ($P_{max}/P_{max}(0), I_{sc}/I_{sc}(0)$, and $V_{oc}/V_{oc}(0)$) of experimental data were interpolated using the expression:
\begin{equation}\label{Eq_fit}
 \left(1-C_1\right)-C\cdot\log_{10}\left[1+\frac{D^{NIEL}(E_d}{D_x}\right]
\end{equation}
\begin{figure}[ht!]
\begin{center}
 \includegraphics[width=1.\textwidth]{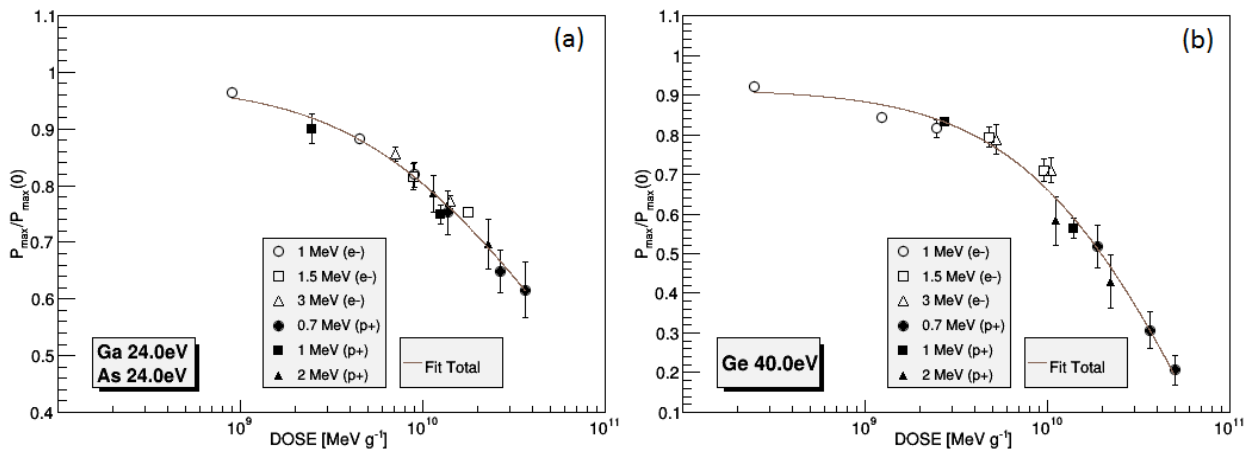}
 \caption{(a) Optimal fit of $P_{max}/P_{max}(0)$ as function of the dose for the 3J solar cell; (b) Optimal fit of $P_{max}/P_{max}(0)$ as function of the dose for single junction bottom cell.}
 \label{fig:Pmax}
\end{center}
\end{figure}
where $C_1$, $C$ and $D_x$ are obtained by a fit to the data in which the NIEL threshold energy, $E_d$, was varied to minimize the differences among electrons and protons data with respect to the corresponding curve obtained from eq. \ref{Eq_fit}.\\
It should also be noted that $C_1$ is only relevant for the bottom cell, while it is negligible in the cases of 3J and GaInP, GaAs single cells. The optimal fit for the TJ cell was obtained using $E_d\approx24$ eV for Ga and As, while we obtained $E_d\approx40$ eV for Ge in the bottom junction (see figure \ref{fig:Pmax}).

\section{Dlts analysis}
In order to perform DLTS (Deep Level Transient Spectroscopy) investigations of deep levels induced by electrons and protons irradiations, mesa-structures of 0.5 mm in diameter have been prepared on top and middle junctions by means of optical photolithography and metal evaporation. Top and middle junctions have exactly the same epitaxial structure of the related component cells.\\
\begin{figure}[ht!]
\begin{center}
 \includegraphics[width=0.8\textwidth]{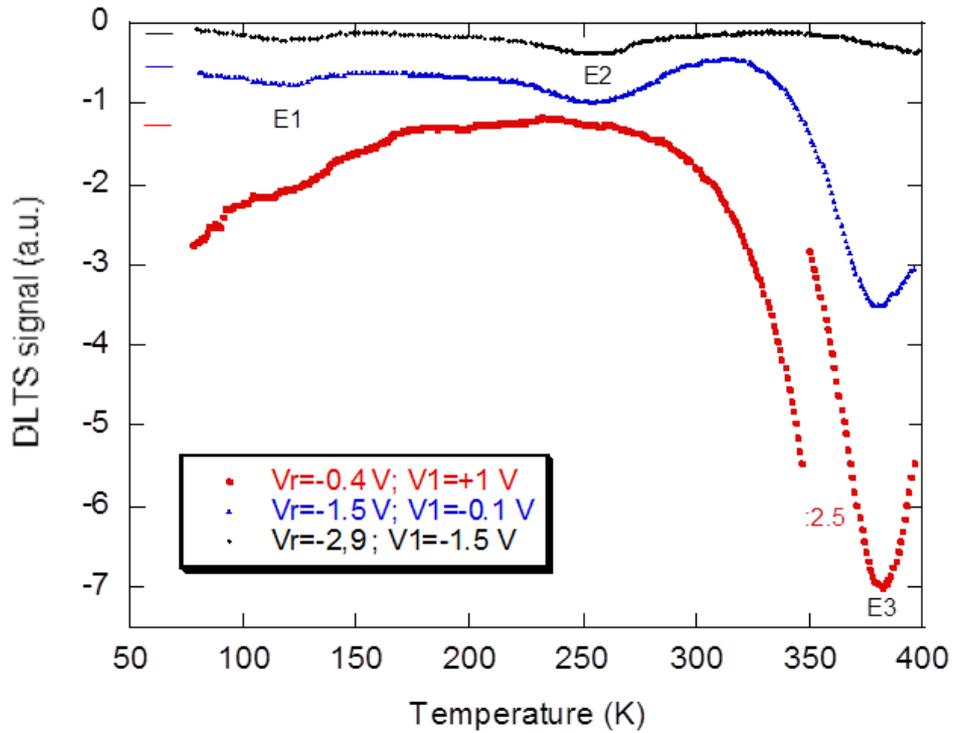}
 \caption{DLTS spectra obtained on a middle junction irradiated  with electrons  (1 MeV at a fluence of $10^{15}$ cm$^{-2}$) using different reverse voltage $V_r$ at fixed pulse amplitude of 1.4 V. Emission rate=46 s$^{-1}$, pulse width=500 ms, V1=pulse voltage.}
 \label{fig:DLTS}
\end{center}
\end{figure}
These samples were irradiated with electrons and protons, simultaneously with the TJ and component solar cells. Fig. \ref{fig:DLTS} shows the DLTS spectra of an electron irradiated middle junction obtained using different reverse voltages $V_r$ (-0.4 V, -1.5 V, and -2.9 V) at fixed pulse amplitude (1.4 V).  The electrons fluence was 10$^{15}$ cm$^{-2}$ and the corresponding NIEL dose $\approx1.1\cdot10^{10}$ MeV/g.\\
With increasing reverse voltage $V_r$ the spectra show the characteristics of regions at increasing distance from the n+/p interface. From the figure it can be seen that the amplitude and hence the trap concentration of the high temperature peak (activation energy=0.71 eV, labelled E3) drastically reduces with the distance from the interface.\\ Considering that a peak with the same characteristics is also present in the non-irradiated samples, E3 is likely to correspond to a defect at the junction interface. For higher reverse biases the DLTS spectra show the presence of at least two levels, labelled E1 (0.21 eV) and E2 (0.45 eV), which are not observed in non-irradiated samples and therefore are attributed to electron irradiation induced defects.  In the spectrum obtained using the lowest reverse bias (-0.4 V) the E1 and E2 peaks are not observable, as they are covered by the presence of a broad DLTS signal most likely due to low energy  interface defects.\\
\begin{figure}[ht!]
\begin{center}
 \includegraphics[width=0.8\textwidth]{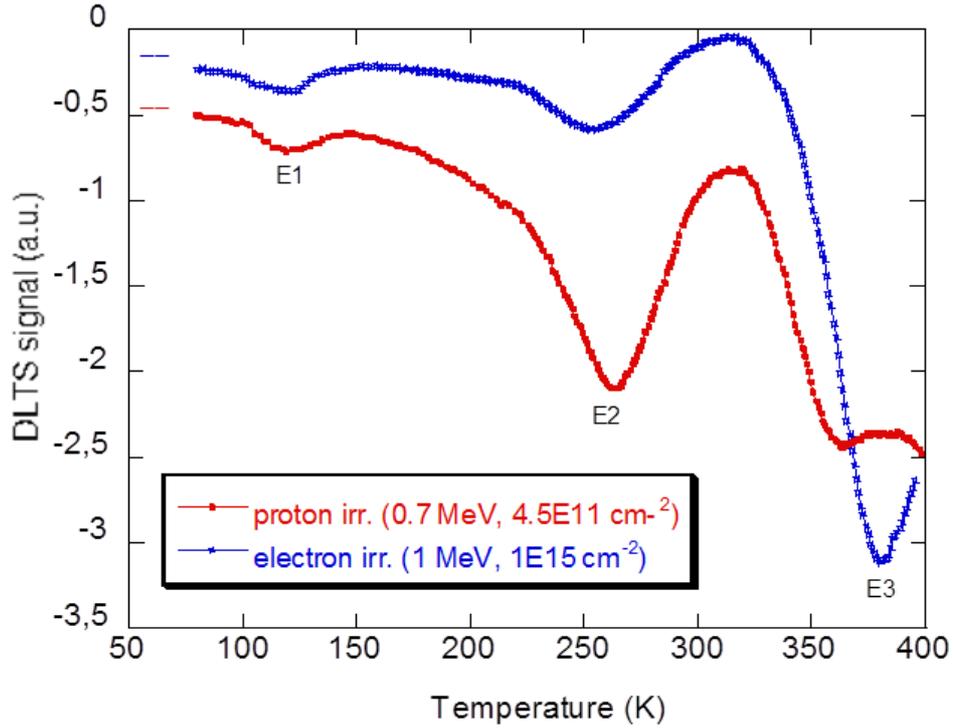}
 \caption{Comparison of the DLTS spectra of two middle junctions irradiated by protons and electrons respectively. Emission rate=46 s$^{-1}$, pulse width=500 ms, reverse voltage $V_r$=-1.5 V, pulse voltage V1=-0.1 V.}
 \label{fig:DLTScomp}
\end{center}
\end{figure}
In Fig. \ref{fig:DLTScomp} the DLTS spectrum of a middle cell diode irradiated with protons  (energy 0.7 MeV and fluence $4.5\cdot10^{11}$ cm$^{-2}$ corresponding to NIEL dose of $3\cdot10^{10}$ MeV g$^{-1}$) is compared to that of a middle cell diode irradiated with electrons (energy 1 MeV and fluence $1\cdot10^{15}$ cm$^{-2}$ corresponding to NIEL dose of 1.1$\cdot10^{10}$ MeV g$^{-1}$). From the analysis of the figure the most important observations are:
\begin{enumerate}
\item two or more DLTS peaks at high temperature are present in the proton irradiated sample, while a single peak E3 is present in the electron irradiated sample.
\item in both electron and proton irradiated samples the peaks E1 and E2 and a broad low temperature shoulder of  peak E2 are present.
\item the ratio of the peak amplitudes E2/E1 is observed to be much larger for the proton irradiated sample than for the electron irradiated one.
\end{enumerate}
Currently further DLTS investigations are being performed on top and middle junctions to obtain a more complete picture of  the dependence of irradiation induced deep levels on electron and proton energy and fluence.

\section{Conclusions and future work}
TJ InGaP/GaAs/Ge solar cells and related component cells have been irradiated with protons and electrons at different energies.\\
The data have been analyzed using the DDD methods by the evaluation of the NIEL. Very peculiar results were obtained for the bottom junction that showed high degradation after proton irradiation.\\
The analysis by DDD requires an additional parameter to take into account, at the same time, the behavior of the bottom junction cell and TJ. However, the parameter C1 is particularly relevant only in the description of dose dependence for the Ge bottom cell.\\
DLTS analyses, carried out on middle junctions, indicate that complex defects are introduced at a different rate for electron and proton irradiations.

\end{document}